# Bulk and contact-sensitized photocarrier generation in single layer TPD devices

Debdutta Ray[1], Meghan P.Patankar[1], Gottfried H.Döhler[2], and K.L.Narasimhan[1,*]

[1] Tata Institute of Fundamental Research, Colaba, Mumbai, 400005, India

[2] Max Planck Research Group, Institute of Optics, Information and Photonics, University Erlangen-Nuremberg, Germany

## Abstract

In this paper, we report on the photoelectronic properties of N,N'-Diphenyl-N,N'-bis(3-methylphenyl)-(1,1'-biphenyl)-4,4'-diamine (TPD) studied in sandwich geometry. In particular, we have obtained from both forward and reverse bias measurements the $\mu\tau$ product for holes in TPD. $\mu$ is the hole mobility and $\tau$ the carrier trapping time. The $\mu\tau$ product is a measure of the electronic quality of the material and allows a quantitative comparison of different samples. We have carried out numerical simulations to understand the photocurrent in these structures. We show that in reverse bias, the photocurrent (PC) is due to bulk. The carrier generation is governed by field assisted exciton dissociation at electric fields greater than $10^6$ V/cm. At lower fields the generation of carriers occurs spontaneously in the bulk of the sample. In forward bias, the photocurrent is due to exciton dissociation at the ITO contact. We also obtain a $\mu\tau$ product for holes from forward bias PC measurements which is in agreement with the value obtained from reverse bias measurements. Based on our experiments, we demonstrate that TPD in a sandwich structure is a good candidate for cheap large area solar blind UV detector arrays.

[*] *Corresponding author, Electronic mail: kln@tifr.res.in*

## 1. Introduction

The discovery of efficient electroluminescence[1] in organic semiconductors has triggered revival of interest in organic semiconductors. This interest has opened up new areas of investigation in optoelectronics based on organic semiconductors. Some examples are OLEDs, field effect transistors, solar cells etc.[2-4] However, electronic transport and optoelectronic properties in these materials are not well understood and are currently the focus of many investigations.[5]

TPD is widely used as a hole transporting layer in bi-layer organic light emitting diodes based on small molecules. DC transport in this material has attracted a great deal of attention. Contacts significantly influence electronic transport in TPD.[5] However, very little is known about the optoelectronic properties of TPD. Yang et. al.[6] have reported on the photoconductivity in TPD and conclude that photocarrier generation in TPD is due to exciton dissociation at the contact. However, no detailed study of the optoelectronic properties of TPD has been reported in the literature. One of the advantages of studying phototransport is that it is possible to get an estimate of the $\mu\tau$ product of the majority carrier where $\mu$ is the carrier mobility and $\tau$ the carrier lifetime.

The $\mu\tau$ product is the figure of merit for a photoconductor and is a measure of sample quality. The $\mu\tau$ product enables a quantitative comparison of different samples of the same material in terms of electronic quality. This provided the motivation for studying photoconductivity in TPD. Recently we had obtained the $\mu\tau$ product of electrons in a widely used electron transporting molecule – tris (8-hydroxyquinoline) aluminum ($Alq_3$) using photoconductivity measurements.[7]

The paper is organized as follows. Section 2 contains a brief description of the experimental set up. Section 3 contains results and discussion. This is divided into three parts. After a brief general discussion of the data in forward and reverse bias (3A) this section is followed by a detailed discussion on photoconductivity in reverse bias (3B). To properly understand transport in reverse bias, we present simulations of phototransport in organic semiconductors followed by a discussion of results of photoconductivity in reverse bias. Section 3C discusses the photoconductivity in forward bias followed by conclusions. Appendix 1 contains details of the simulations used in section 3B.

**2. Experiment**

The TPD devices (sandwich geometry ITO/TPD/Al/LiF) were prepared by evaporating TPD on pre-patterned ITO coated glass substrates in vacuum at a pressure of $8 \times 10^{-7}$ Torr. The top contact was semitransparent Al. The devices were finally encapsulated in-situ with a 1000Å LiF layer. The thickness of the TPD samples was varied from 500-3000Å. The thickness was measured in-situ using a quartz crystal monitor and confirmed independently by optical absorption and capacitance measurements. The photocurrent was measured by illuminating the devices through the (bottom) ITO or (top) Al electrode. The light sources used were a 100W Xenon lamp coupled to a monochromator and a 360nm LED. The light was chopped using a mechanical chopper.

All measurements were made with the sample in vacuum. The photocurrent (PC) defined as $J_{PC} = J_{Light} - J_{Dark}$, was measured using a lockin amplifier. Details of sample preparation and measurements have been described earlier.[7]

**3A. Results and Discussion**

Figure 1 shows the PC of 2000Å and 1000 Å thick TPD samples as a function of applied electric field for 360 nm (hυ=3.44 eV) illumination. The inset shows the schematic energy level diagrams in forward and reverse bias conditions for the sample. In forward bias (i.e. ITO positively biased), the photocurrent initially increases strongly with bias and then tends to saturate. The PC is strongly asymmetric – it is much larger in forward bias than in reverse bias. We also see from the figure that the magnitude of the saturation photocurrent in forward bias depends strongly on the following:

a) The thickness of the sample.

b) Whether the sample is illuminated through the bottom ITO or top aluminium electrode. The PC is much larger for illumination through the bottom (ITO) electrode.

Figure 2 shows the spectral response of the sample in reverse and forward bias respectively for a 2000 Å thick sample of TPD. In this experiment light is incident through the top aluminum electrode. The optical absorption spectrum of the sample (TPD) is also shown for comparison. The spectral response of the photocurrent is very different for the two cases. In forward bias, the PC is maximum (minimum) when the absorption is minimum (maximum) – the spectral response is antibatic with respect to the absorption spectrum. In reverse bias the photocurrent spectral response is symbatic with the optical absorption. We now try to understand these results.

Optical excitation in organic semiconductors is primarily excitonic in nature.[8] Free carrier production is associated with exciton dissociation. This can take place either in the bulk of the sample or at the electrodes. We now briefly discuss these two cases.

We first discuss the case of exciton dissociation at the electrodes giving rise to PC. In this case, excitons generated within a diffusion length of the electrode diffuse to and dissociate at the electrode into electrons and holes.[9,10] One of the carriers will recombine with the image charge at the electrode and the other carrier drifts under the influence of the external electric field giving rise to photocurrent. Only these excitons contribute to the PC. It is well known that holes dominate electronic transport in TPD.[11,12] In forward bias, for light incident through the top electrode (Al), the PC is a maximum when the optical absorption is a minimum (corresponding to maximum light reaching the ITO surface) and vice versa. This is consistent with the data in fig.1 where in forward bias (for a 1000 Å thick sample) the saturated PC is almost an order of magnitude larger for light incident through the ITO contact when compared with illumination though the aluminium contact. We hence conclude that in forward bias, the PC is primarily due to exciton dissociation at the ITO surface. This will be discussed in greater detail later in this paper.

If, on the other hand, the PC is a bulk process, then one expects that for uniformly absorbed light, the magnitude of the PC should be the same whether the sample is illuminated through the top or bottom electrode.[7] The criterion of uniform illumination implies thin samples ($\alpha L \ll 1$). The optical absorption co-efficient ($\alpha$) for TPD, at 360 nm, is about $1.8x10^5$ $cm^{-1}$. The 1000 Å thick sample does not quite satisfy the criterion for uniform absorption when illuminated at 360nm. However we see from fig.1 that the photocurrents for a 1000 Å thick sample in reverse bias when illuminated through either ITO or the Al electrodes are almost equal and differ by only a factor of two. This is so even though the light is attenuated by almost an order of magnitude while traversing the length of the film. We contrast this with the situation in forward bias (where the PC is primarily due to exciton dissociation at the ITO electrode) - the PC is smaller by an order of magnitude for illumination through the Al contact as compared with illumination through the ITO contact. (Holes created in the vicinity of the ITO electrode by exciton dissociation recombine at the ITO electrode and do not contribute to the photocurrent in reverse bias.) We hence conclude that PC in reverse bias is a bulk effect. We now present simulations of the photocurrent in sandwich structures to understand the bulk PC in reverse bias. This is then followed by a detailed discussion of the PC in forward bias.

**3B. Numerical Simulation Results (Reverse Bias)**

TPD is known to be a hole transporting layer. The barrier to electron injection (Al) is large (~ 1eV) while the barrier to hole injection (ITO) is around 0.2 to 0.4eV. We are dealing with a system which has non-injecting contacts in reverse bias. The PC of such systems has been studied earlier.[13-16] Insight into the PC process can be obtained by simulation studies. The PC can be obtained by a simultaneous solution of the continuity and Poisson equations and is described in detail in Appendix 1. We use the results of the simulations to understand the PC in organic semiconductors. In this simulation $b=\mu_p\tau_p/\mu_n\tau_n$ where $\mu_p$ and $\mu_n$ are the hole and electron mobilities and $\tau_p$ and $\tau_n$ are the trapping times for holes and electrons respectively. The sample thickness (L) has been taken to be 1000Å. Figure 3 shows the carrier recombination as a function of position in the sample for uniformly absorbed light for different values of b for a constant value of photocurrent. We see from the figure that the recombination profile in the sample broadly breaks up into three regions. In the vicinity of the contacts, the recombination of carriers is small i.e. R<<G where R and G are the recombination and free carrier generation rates respectively. The width of these regions is given by $l_p$ and $l_n$ respectively. These two regions are separated by a region of width $l_R$ where R = G. We can hence think of the sample as a stationary region ($l_R$) in series with current generation regions ($l_p$ and $l_n$) respectively.[15] $l_p$ and $l_n$ are the collection lengths and are given by $\mu_p\tau_pF$ and $\mu_n\tau_nF$ respectively where F is the electric field. In the figure, $l_p$ and $l_n$ corresponds to the collection length for holes (near x=0) and for electrons (near x=L) respectively. As b increases, we see from the figure that the electron collection length $l_n$ decreases. The photocurrent can be written as,[14]

$$J = q\gamma\eta(l_p + l_n)[1-\exp(-L/(l_p + l_n))] \qquad (1)$$

where η is the internal quantum efficiency for free carrier production and γ is the generation rate. For $(l_p + l_n) << L$, Eq. 1 can be approximated to,

$$J = q\gamma\eta(l_p + l_n) = q\gamma\eta F(\mu_p\tau_p + \mu_n\tau_n) \qquad (2)$$

We see from equation 2 that if η is independent of the electric field, the PC is dominated by the carrier with the larger collection length. This is in agreement with earlier calculations of the PC in insulating systems.[13-16] As the electric field increases, regions $l_p$ and $l_n$ grow at the expense of $l_R$ till $l_R$ is reduced to a point-which represents complete carrier collection. In such a case, the PC will be independent of bias and can be written as,

$$J_{sat} = q\gamma\eta L \qquad (3)$$

This has been recently demonstrated in the case of $Alq_3$.[7]

In organic semiconductors, the mobility also depends on the electric field[11,12] in many systems and can be written as,

$$\mu_j = \mu_{0j} \exp(\beta_j \sqrt{F}) \qquad (4)$$

where the suffix j is used to denote electrons (holes).The expression for the photocurrent (eq. 2) now has an electric field dependent carrier mobility given by eq. 4. As an example we choose the field dependent carrier mobilities corresponding to that of $Alq_3$.[7,17] Using the method outlined in Appendix 1, we calculate the photocurrent–voltage characteristics. The carrier generation rates were taken to be around $2\text{x}10^{18}$ $cm^{-3}s^{-1}$ which are similar to that used in the experiments. For these generation rates space charge effects were negligible. Hence the electric field was found to be constant in the sample. This is in agreement with experiment where the photocapacitance and dark capacitance were the same in these samples - consistent with the existence of a uniform electric field in the sample under illumination.[18] Figure 4 is a simulated plot of log ($J_{pc}$ / F) vs $F^{1/2}$. We see from the figure that the slope of the curve is determined by $\beta_e$ -characteristic of electrons – which is the dominant carrier in $Alq_3$. These results are in good agreement with experimental results reported elsewhere.[7]

The results summarized above are valid for light uniformly absorbed in the sample. We now discuss the case for non-uniformly absorbed light. In this case, the generation (and the recombination) is a strong function of position in the sample. We see from figure 1 that in reverse bias, the magnitude of the current for a 2000 Å thick sample depends strongly on whether the light is incident through the anode (Al) or the cathode (ITO). This result has a simple physical explanation.[14] When strongly absorbed light is illuminated through the top Al contact (anode), the virtual cathode moves into the bulk of the sample. To a first approximation we can assume that light is uniformly absorbed in the region between the anode and the virtual cathode. Beyond this region, the recombination becomes negligible as free carrier generation is also reduced. $l_p$ is near the top contact and $l_n$ in the vicinity of the virtual cathode. Holes generated in both $l_p$ and $l_n$ will transit to the cathode (ITO) and contribute to the photocurrent. On the other hand when strongly absorbed light is incident through the ITO contact (negative bias), the virtual anode (Al) moves into the bulk of the material. Only holes generated in $l_p$ contribute to the photocurrent. (Holes generated in $l_n$ adjacent to the ITO electrode do not contribute to the PC.) The light reaching $l_p$ (adjacent to the virtual anode) is attenuated due to absorption. The total transit length of the hole is less than L. Both these factors contribute to a reduction in the photocurrent. This qualitatively accounts for the asymmetry in the PC- Voltage characteristics for strongly absorbed light incident through the top (bottom) electrode. We hence conclude that this criterion (viz. the PC magnitude is independent of the light incidence direction) that is commonly used for determining whether the PC is a bulk or exciton dissociation at the contact is strictly valid only for uniformly absorbed light i.e. when $\alpha L << 1$.

Figure 5 shows the PC response in reverse bias for a 2000Å sample. We see from the figure that the PC response varies linearly with electric field up to $8x10^5$ V/cm. In this figure, we have corrected for the built in voltage of the sample. The linear increase of PC with electric field is in agreement with eq. 2. In this range, this implies that both $\eta$and $\mu$ are independent of the electric field. Using eq.2, we estimate the low field $\eta\mu\tau$ product for holes to be $4.3x10^{-14}$ $cm^2/V$ in our samples. The $\eta\mu\tau$ product is a figure of merit for the photoconductor and can be used to quantitatively compare different samples of TPD. Attempts to measure $\eta$ from total collection measurements were not successful – as the PC increases nonlinearly with bias at higher voltages. Even at a field of $2x10^6$ V/cm, for a 500 Å thick film, the PC did not reach saturation. This suggests that either $\eta$ or (and) $\mu$ are field dependent quantities. The internal quantum efficiency $\eta$ can increase with electric field due to an Onsager-like process.[19,20] To estimate the field dependence of $\eta$ we measured (in reverse bias) the electric field quenching of photoluminescence (PL). The PL quenching efficiency can be written as,

$$\eta(F) = \frac{PL(0) - PL(F)}{PL(0)} \tag{5}$$

where PL(0) and PL(F) are the PL at zero applied field and at field F respectively.

Figure 6 shows the PL quenching efficiency as a function of applied field for a TPD sample. We see that at a field of $2x10^6$ V/cm, the quenching efficiency is 0.5%. From the log[$\eta$(F)] vs log(F) plot (fig. 6, inset), we find that $\eta$(F) is proportional to $F^{2.5}$.

The electric field quenching of excitons should increase the free carrier generation rate. To see if the field quenching of excitons can account for the non-linearity of the PC with bias at large reverse bias, we assumed that at the highest field ($2x10^6$ V/cm), the PC was determined by the field quenching of excitons. Figure 7 is a plot of the PC data and the normalized PL quenching data. The PL quenching data was normalized to coincide with the PC data at the highest electric field.

For fields between $8x10^5$ V/cm to $2x10^6$ V/cm, $\eta(F)$ closely follows the PC data. This implies that for F > $8x10^5$ V/cm the increase of PC with field is primarily due to increase in generation rate with electric field. There is total collection of carriers in this regime. Below $8x10^5$ V/cm., the carrier generation efficiency $\eta$ is field independent and has a different physical origin. This could be due to weak C-T transitions which are enhanced in the vicinity of the strong absorption[21] and are directly measured in a PC measurement. Using the data of Fig.6 and Fig.7 we estimate the low field value of $\eta$ to be about $10^{-3}$. This allows us to obtain the $\mu\tau$ product for TPD which we estimate to be about $4.3x10^{-11}$ $cm^2/V$.

Using the results discussed here, we demonstrate in Figure 8 that ITO/TPD/Al device in reverse bias can be used as a solar blind UV detector with a response of 1 mA/W. Although this is two orders of magnitude smaller in sensitivity than a GaN based device,[22] the ease and low cost of manufacture and the possibility of low cost large area arrays is in its favour for specific niche applications.

**3C. Photocurrent in Forward Bias**

We now briefly discuss the photocurrent in forward bias (FB). We see in Fig.1 that the photocurrent - voltage characteristics is very asymmetric between forward and reverse bias. At low field ($<10^5$ V/cm), the PC in FB rises exponentially with electric field and is very much larger in FB than in RB. This asymmetry can be a consequence of photoconductive gain or a possibility that the PC is due to exciton dissociation at the electrode (ITO for TPD). The gain is defined as,

$$Gain = \frac{J_{pc,forward}}{J_{pc,reverse}}. \tag{6}$$

The built-in voltage is taken into account in computing the gain. For light through ITO the gain is about 100 at low forward bias. For illumination through Al, the gain depends on the thickness of the film and varies between 2 to 10 for the samples discussed here. If $t_{tr}$ is the transit time through the sample the gain can also be written as;

$$Gain = \tau / t_{tr} = (l_p + l_n)/L \tag{7}$$

From the earlier discussion, we know that $(l_p+l_n) << L$ at a field of $10^5$ V/cm. Hence photoconductive gain is not responsible for the asymmetry in the photocurrent-voltage characteristics. Asymmetry in the photocurrent-voltage characteristics can also arise if the PC is due to exciton quenching at the electrode in FB. This is consistent with our earlier identification from spectral response measurements that exciton dissociation at the ITO electrode determines the PC in forward bias. The photocurrent due to exciton dissociation for illumination through the bottom (ITO) can be written as,[23]

$$J_{bottom} = \frac{\alpha \phi Q_0 e}{(1/L_D) + \alpha} \tag{8}$$

where $Q_0$ is the exciton quenching efficiency at the electrode, $\varphi$ the incident photon flux density and $L_D$ the exciton diffusion length. For $(1/L_D) >> \alpha$, the $Q_0 L_D$ product is estimated from the saturated photocurrent to be $1.2 \times 10^{-7}$ cm.

We have also independently measured the exciton quenching at the ITO electrode by measuring the PL yield for samples of different thickness.[24] These experiments were carried out by simultaneous deposition of a series of TPD films of different thickness on quartz and ITO coated glass substrates. The thickness of the TPD films was varied from 5 nm to 20 nm. The PL of the samples deposited on ITO was normalized using the corresponding sample (of the same TPD thickness) on quartz substrates. Beyond 10 nm, the PL of the sample on ITO and quartz were similar. At lower thickness, the PL of films deposited on ITO was quenched with respect to that deposited on the quartz substrate. Fig 9 shows the quenching as a function of film thickness. The quenching efficiency for a film with thickness L can be written as,[25]

$$Q = Q_0 \frac{\left[\alpha^2 L_D{}^2 + \alpha L_D \tanh\left(L / L_D\right)\right]\exp(-\alpha L) - \alpha^2 L_D{}^2 \left[\cosh\left(L / L_D\right)\right]^{-1}}{\left(1 - \alpha^2 L_D{}^2\right)\left(1 - \exp(-\alpha L)\right)} \qquad (9)$$

Using eq.9 to fit the data in fig.9, we obtain values for $Q_0$ and $L_D$ to be 0.3 and 50 Å respectively. This is in very good agreement with the $Q_0L_D$ product obtained from the current measurements (eq.8). Similar experiments with quartz / Al / TPD structures (to estimate the exciton quenching by the Al electrode) showed that down to the lowest thickness, there was no detectable quenching of the excitons. Using the value of $Q_0$ and $L_D$ we calculate the spectral response of PC for a 2000Å sample when illuminated through the top Al electrode. The current in this case can be written as,[23]

$$J_{top} = \frac{\alpha \phi Q_0 e}{(1 / L_D) - \alpha} \exp(-\alpha L) \qquad (10)$$

Fig.10 shows the calculated spectral response for a 2000Å sample when illuminated through the top electrode (Al) in forward bias to be in good agreement with the experiment.

We briefly comment on the voltage dependence of PC in forward bias. In forward bias the PC varies strongly with field to about $10^5$ V/cm. Clearly this is related to collection efficiency. Holes generated at the ITO interface drift in the applied electric field to the cathode. If there are deep traps with a trapping time $\tau$, then the photocurrent can be written as,[26]

$$J = J_{sat} \exp(-t/\tau) \qquad (11)$$

where $t = t_{tr} = L/\mu F$ is the transit time through the sample. Fig. 11 is a semi-log plot of the photocurrent as a function of 1/F for a 1000A thick film. From the slope of the $\log(J_{pc})$ vs (1/F) we obtain the value of $\mu\tau$ product to be $6.8\times10^{-11}$ $cm^2/V$. This is consistent with the value for the $\mu\tau$ product obtained from reverse bias measurements in the earlier section.

## 4. Conclusion

In conclusion, we have studied the photoconductivity in TPD in a sandwich structure. In reverse bias, the PC is a bulk effect. At low electric fields ($F<8.10^5$ V/cm), the quantum efficiency of free carrier generation is independent of the electric field and is about 0.1%. At higher fields, the generation efficiency increases due to electric field dissociation of excitons in an Onsager like process. We have obtained a value for the $\mu\tau$ product to be about $4.3\times10^{-11}$ $cm^2/V$. This is an important quantity as it can be used as a quantitative measure of the purity and quality of samples. We also show that TPD in a sandwich structure is a good solar blind UV detector.

In forward bias, the photocurrent is due primarily to exciton dissociation at the ITO electrode. From PL quenching measurements, we have measured the quenching efficiency and the exciton diffusion length in TPD. We have also obtained the $\mu\tau$ product from collection efficiency in forward bias measurements and show that it is consistent with estimates obtained from reverse bias measurements.

We have presented simulation of the PC in sandwich structures of organic semiconductors to understand the photoconductivity in these systems.

## 5. Acknowledgments

It is a great pleasure to thank Professor N. Periasamy for many useful discussions.

## Appendix 1

The photocurrent can be obtained by solving the Poisson and continuity equation simultaneously. In sandwich geometry with non-injecting contacts the photocurrent can be written as:

$$j = j_p(x) + j_n(x) \qquad \text{(A1-1)}$$

$$j_p(x) = e\mu_p p(x)F - \mu_p kT(dp/dx) \qquad \text{(A1-2a)}$$

$$j_n(x) = e\mu_n n(x)F + \mu_n kT(dn/dx) \qquad \text{(A1-2b)}$$

$$\frac{dj_p(x)}{dx} = e(G(x) - R(x)) = -\frac{dj_n(x)}{dx} \qquad \text{(A1-2c)}$$

where $j_n(x)$ and $j_p(x)$ are the electron and hole current respectively. The electric field is given by F. The diffusion current has been neglected in this case. The carrier generation takes place in the bulk of the sample. The spatial free carrier generation profile is given by:

$$G(x) = I(x)\alpha\,, \qquad \text{(A1-3a)}$$

where, I(x) is the intensity (incident through anode) and is given by;

$$I(x) = I_0 \exp(-\alpha x) \qquad \text{(A1-3b)}$$

The recombination can be written[15] as;

$$R(x) = \frac{n(x)p(x)}{n(x)\tau_p + p(x)\tau_n} \qquad \text{(A1-4)}$$

The electric field, given by the Poisson equation is;

$$\frac{dF}{dx} = \frac{e}{\varepsilon\varepsilon_0}(p(x) - n(x)) \qquad \text{(A1-5)}$$

Using eqs. A1-1 to A1-5, the spatial gradient of carriers can be written as;

$$\frac{dp}{dx} = \frac{1}{F}\left[-\left(\frac{e}{\varepsilon\varepsilon_0}\right)p(x)(p(x)-n(x)) + \left(\frac{1}{\mu_p}\right)(G(x)-R(x))\right] \quad \text{(A1-6a)}$$

$$\frac{dn}{dx} = \frac{1}{F}\left[-\left(\frac{e}{\varepsilon\varepsilon_0}\right)n(x)(p(x)-n(x)) - \left(\frac{1}{\mu_n}\right)(G(x)-R(x))\right] \quad \text{(A1-6b)}$$

The boundary conditions at x=0 (anode) is as follows;

$$p(0) = 0,\ j_p(0) = 0\, and\, j_n(0) = j \quad \text{(A1-7)}$$

The electric field is taken as the parameter for the simulation. The simulation starts by assuming a value of F. Using eq. A1-1, A1-2 and A1-7, the value of n(0) is calculated. n(x) and p(x) are found by iteration and the simulation stops when the boundary condition n(L)=0 is satisfied.

**Figure captions**

**Fig 1.** A plot of photocurrent as a function of electric field for 1000 Å and 2000 Å thick samples when illuminated through; 1) ITO (◊:1000 Å and Δ:2000 Å), 2) Al (o:1000 Å and □:2000 Å). The energy of illumination was 3.44 eV. The insets show a schematic energy level diagram under forward and reverse bias conditions.

**Fig. 2.** A plot of the spectral response for a 2000 Å thick TPD film in reverse (Δ) and forward (o) bias when illuminated through Al. The absorbance of the sample (in a.u.) is also plotted (—) for comparison.

**Fig. 3.** A plot of the simulated spatial profile of recombination for the following cases: b=1(—), b=10(---) and b=100(–•–). The photocurrent is the same for all three cases. The value of b is varied by changing the $\mu_n\tau_n$ product while keeping $\mu_p\tau_p$ the same. The anode corresponds to x=0. The regions $l_p$, $l_n$ and $l_R$ are shown schematically for the b=1 case.

**Fig. 4.** A plot of the simulated photocurrent for field dependent mobility and its fit. The parameters used in the simulation are as follows:[7,17] $\mu_{0e} = 4.85\text{x}10^{-7}$ cm$^2$/Vs, $\mu_{0h} = 6\text{x}10^{-11}$ cm$^2$/Vs, and $\beta_e = 4.5\text{x}10^{-3}$ (cm/V)$^{1/2}$ and $\beta_h = 9\text{x}10^{-3}$ (cm/V)$^{1/2}$. The slope is found to be $4.2\text{x}10^{-3}$ (cm/V)$^{1/2}$ and equal to the $\beta$ of faster of the carriers, here electrons.

**Fig. 5.** A plot of photocurrent density as a function of electric field in low reverse bias for a 2000 Å thick sample. The electric field is corrected for the built-in field.

**Fig. 6.** A plot of electric field dependent PL quenching efficiency. The inset is the same data in log scale.

**Fig. 7.** A plot of the photocurrent (o) and the electric field dependent carrier generation efficiency (Δ). The data has been normalized at the common highest electric field.

**Fig. 8.** A plot of the photocurrent response of a 750 Å TPD film in reverse bias showing it can be operated as a solar blind UV detector. The device is biased at -10V and has dark currents of around 10nA/cm$^2$.

**Fig. 9.** A plot of PL quenching efficiency of TPD films on ITO substrate as a function of film thickness. A fit of the data (using eq.9) is also shown.

**Fig. 10.** A plot of the spectral response of a 2000Å thick film in forward bias (o) when illuminated through Al. The calculated photocurrent is also plotted (—) for comparison.

**Fig.11.** A plot of photocurrent density as a function of 1/F.

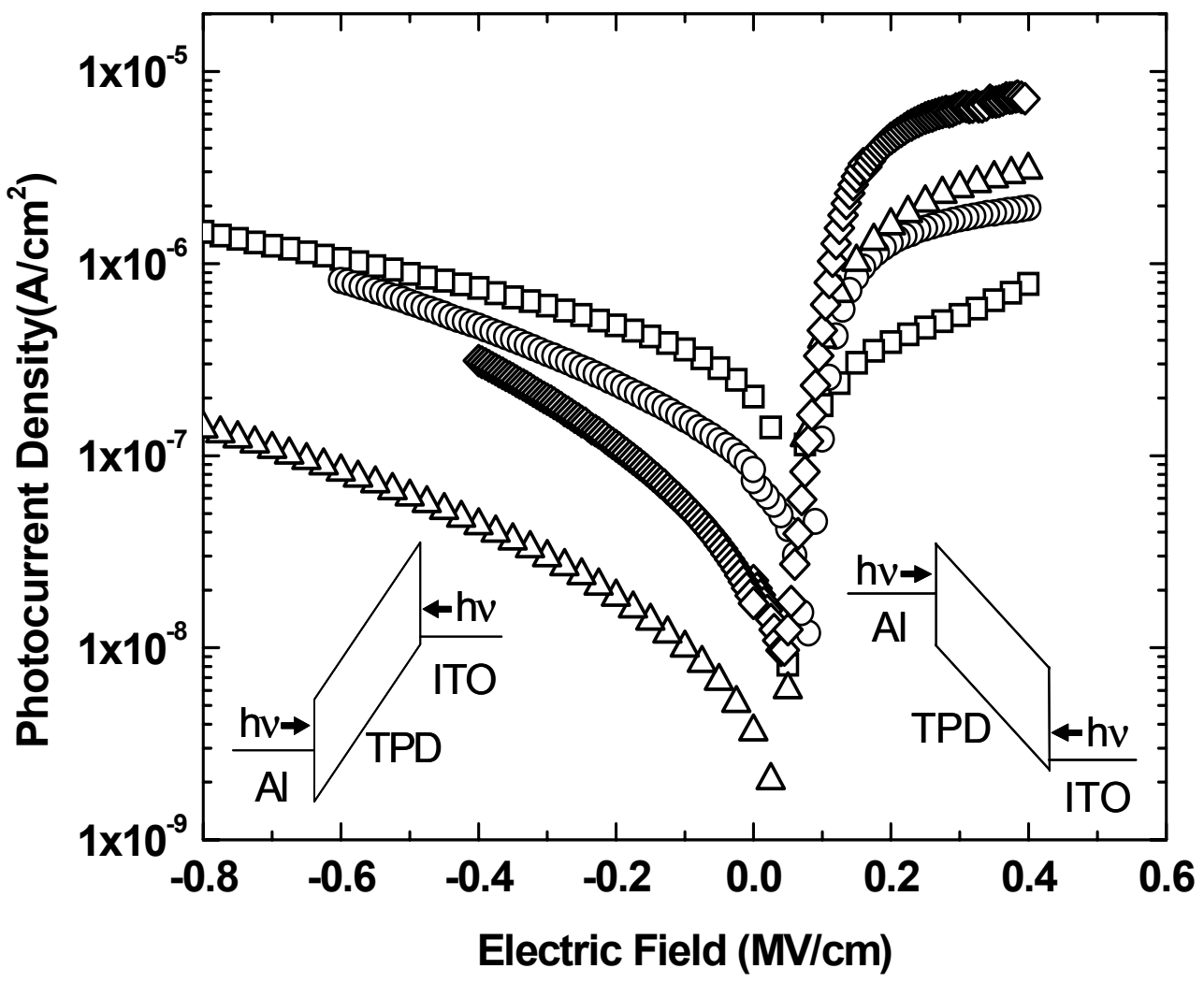

**Fig. 1.** Debdutta Ray et al.

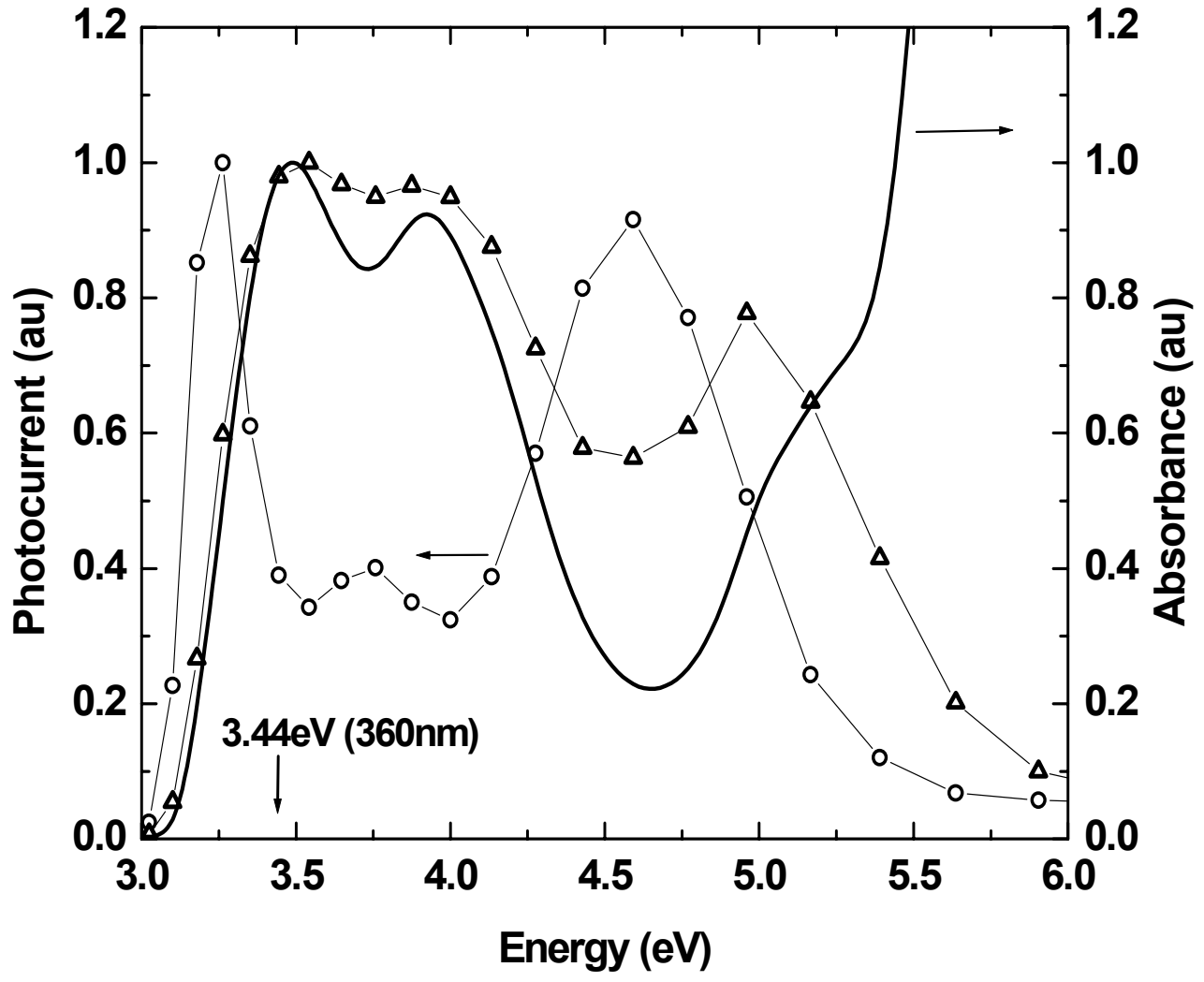

**Fig. 2.** Debdutta Ray et al.

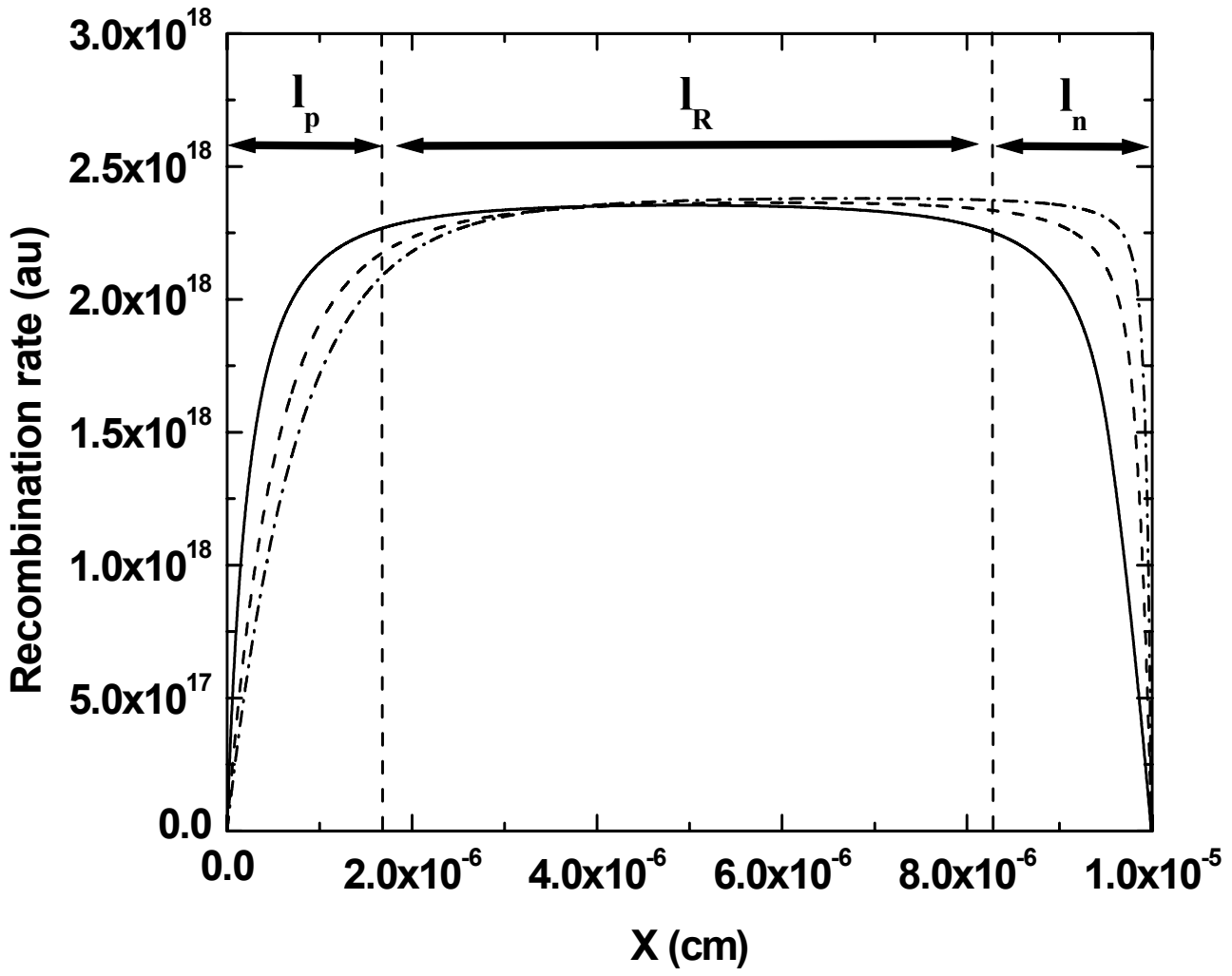

**Fig. 3.** Debdutta Ray et al.

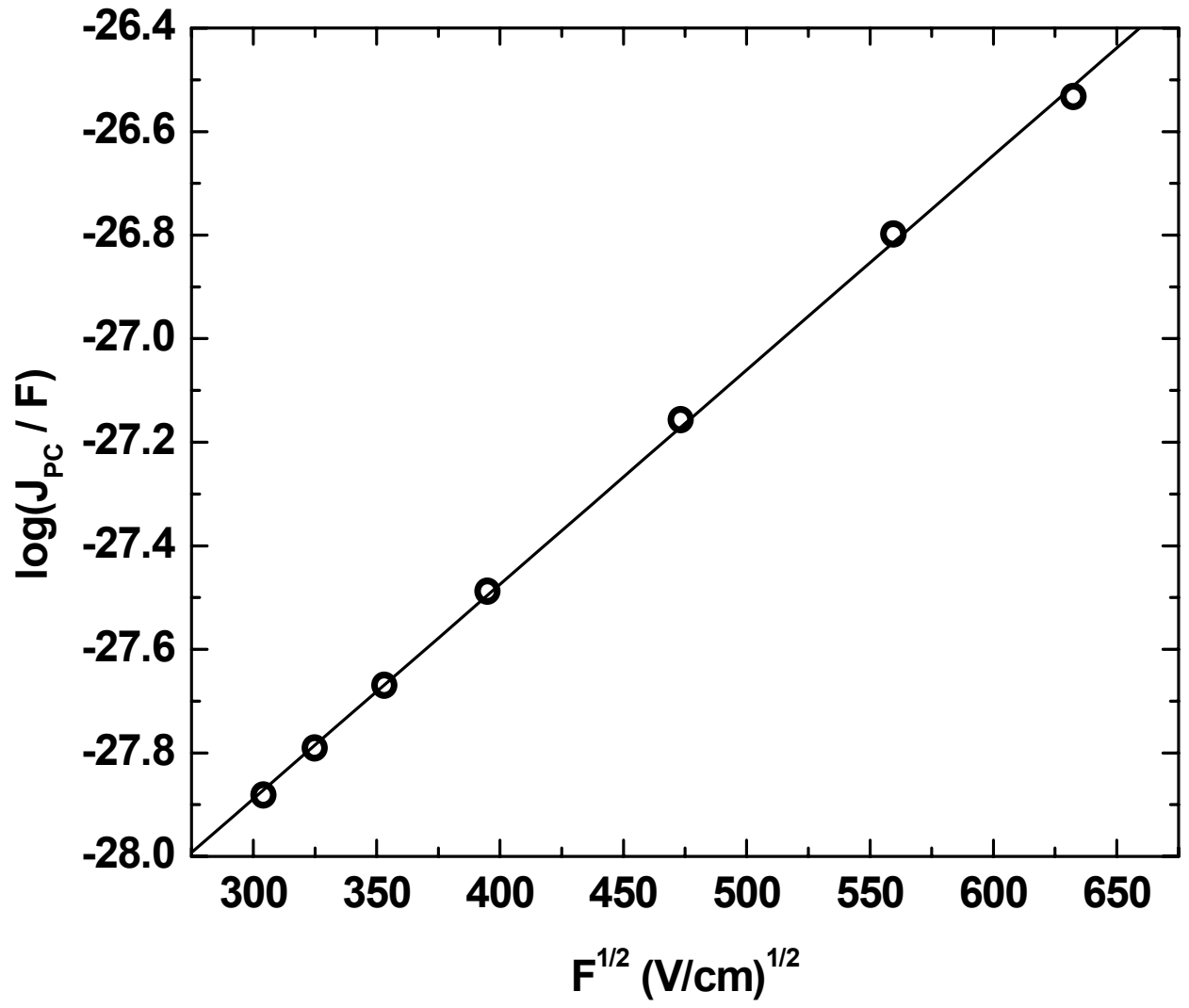

**Fig. 4.** Debdutta Ray et al.

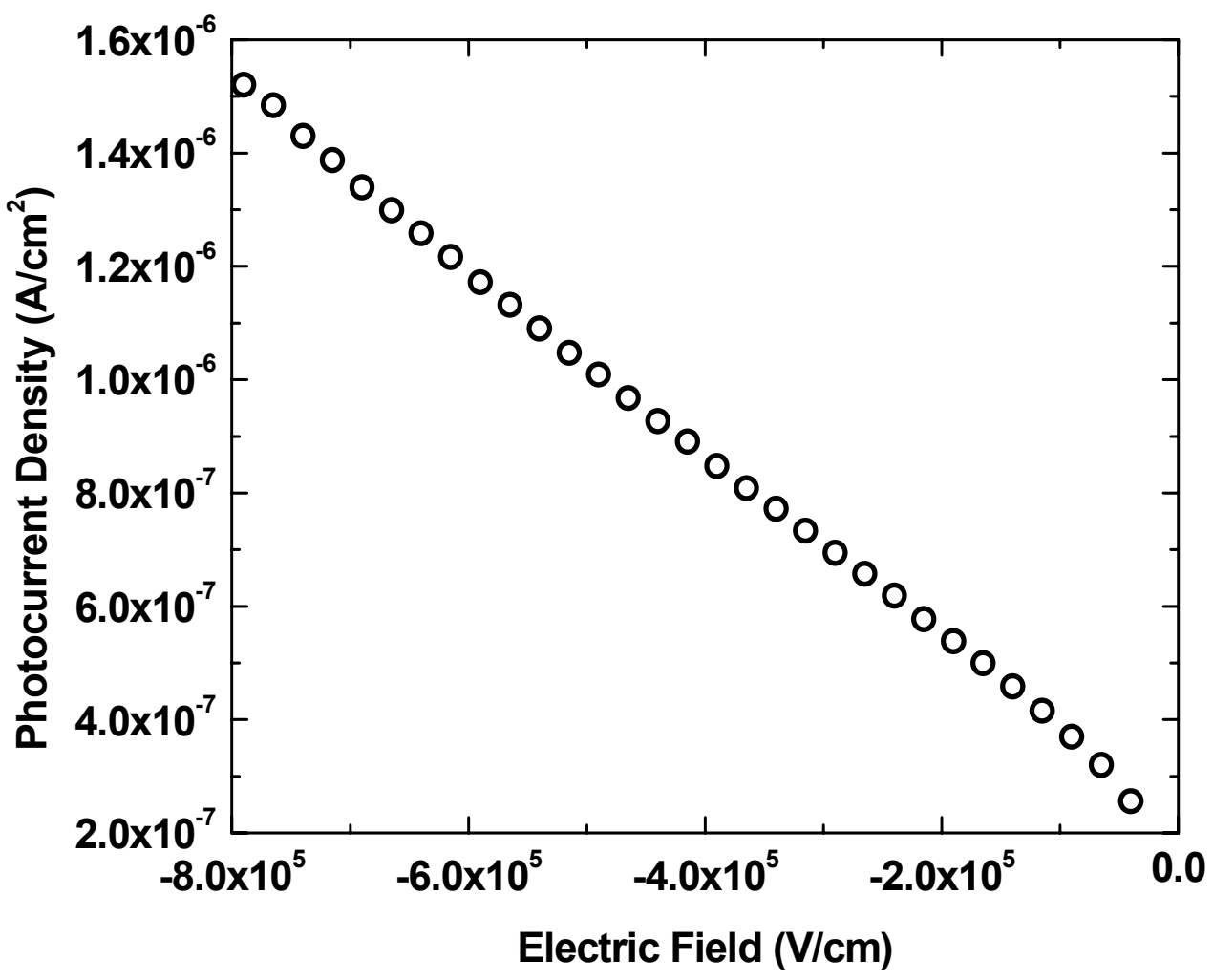

**Fig. 5.** Debdutta Ray et al.

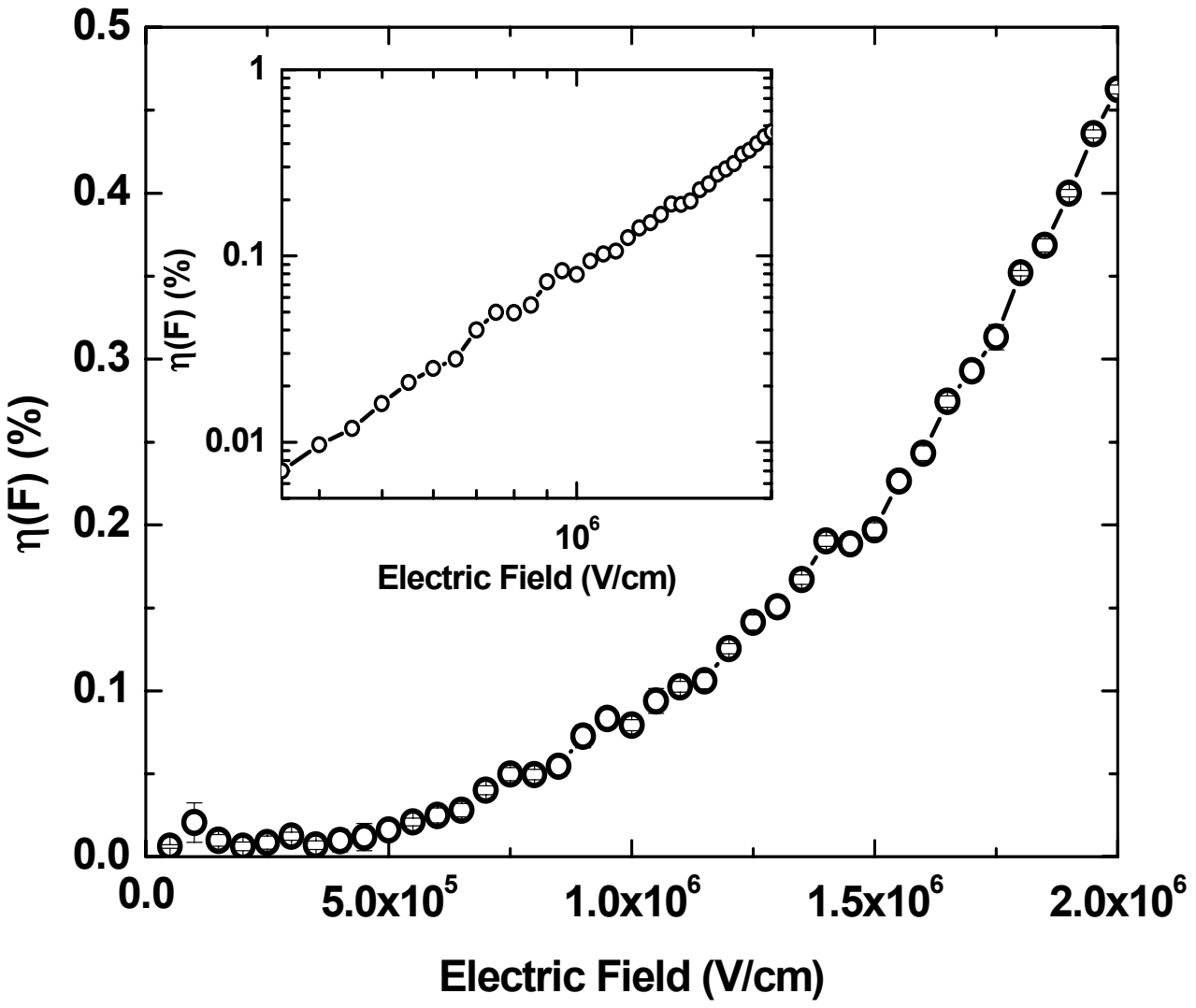

**Fig. 6.** Debdutta Ray et al.

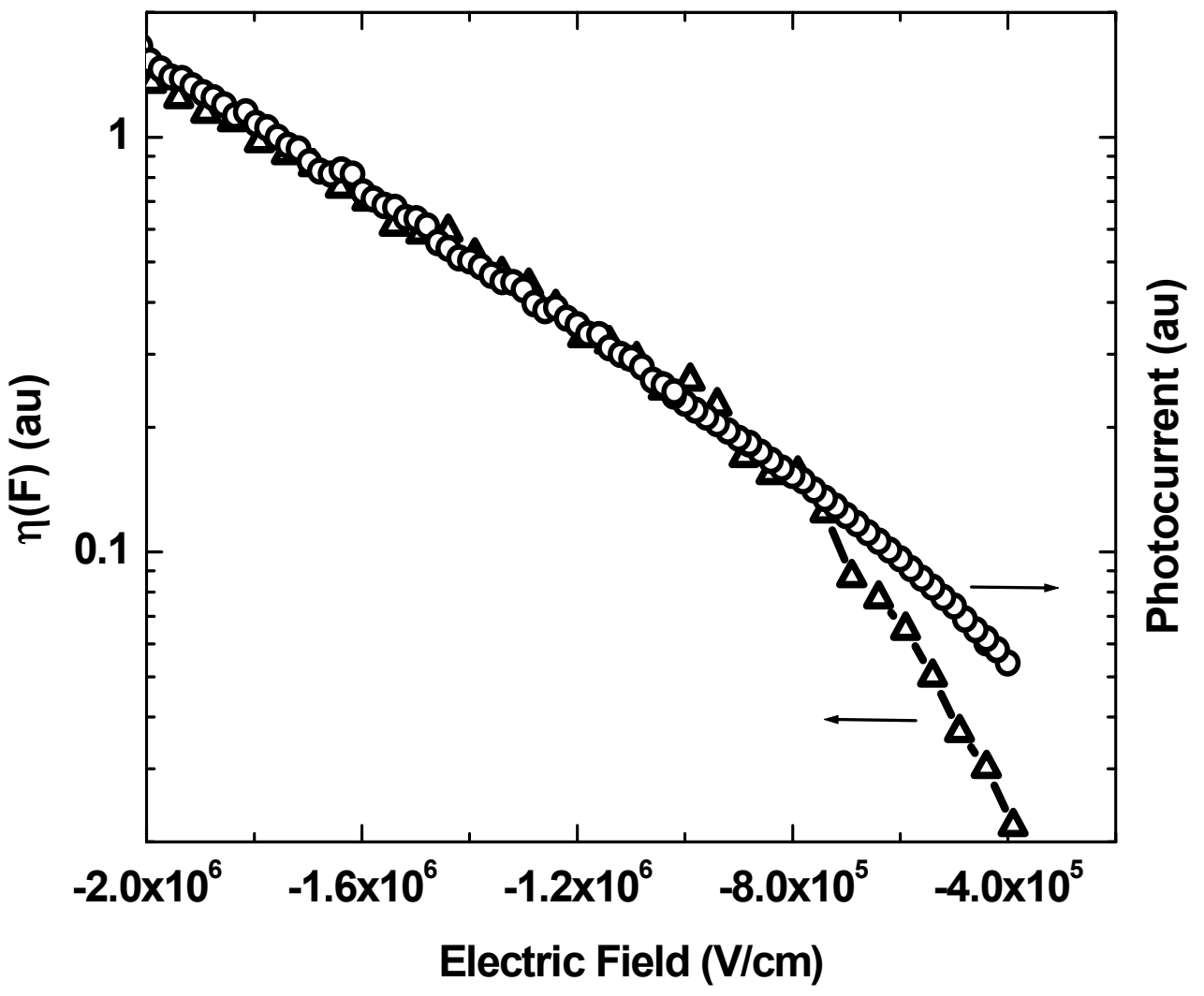

**Fig. 7.** Debdutta Ray et al.

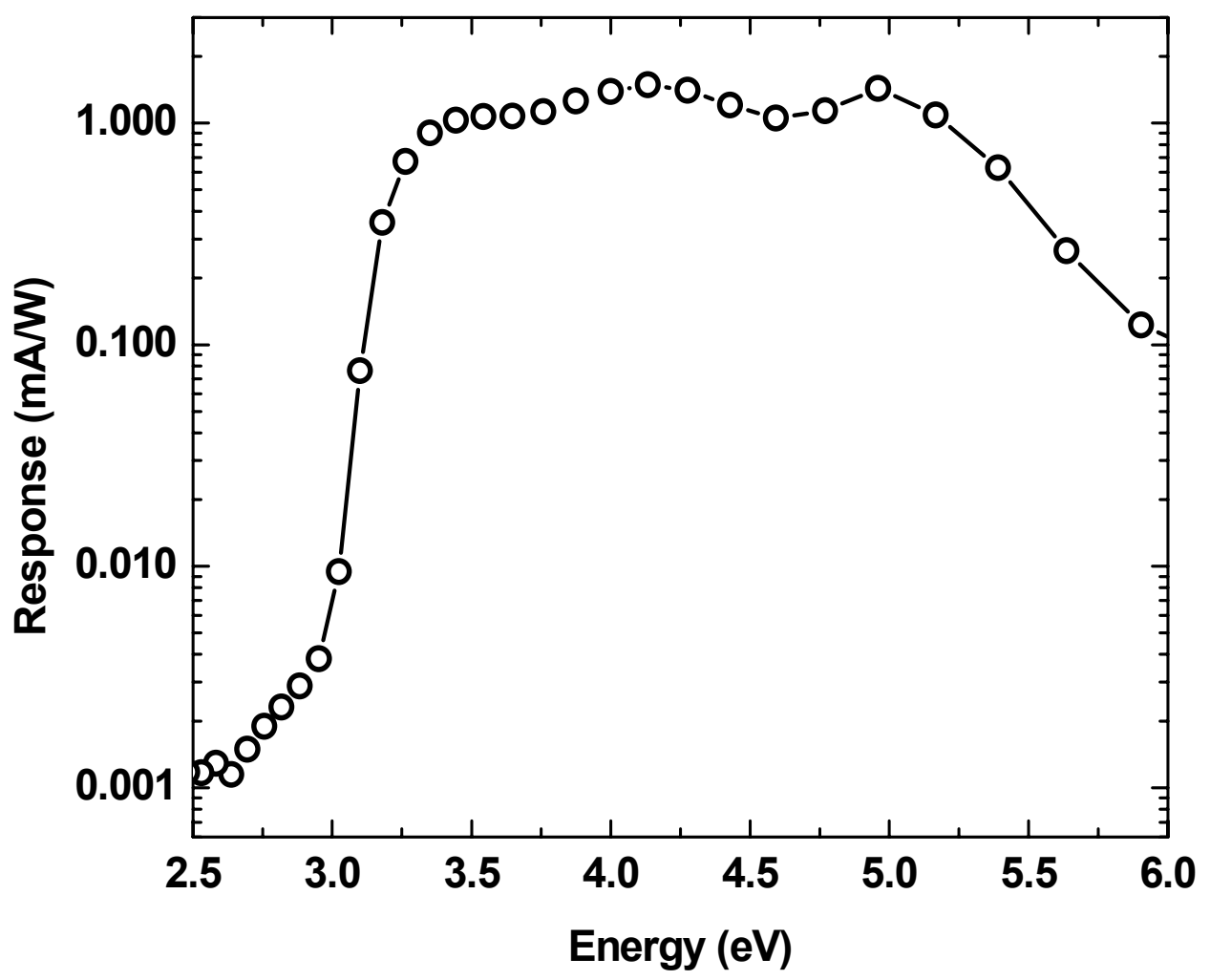

**Fig. 8.** Debdutta Ray et al.

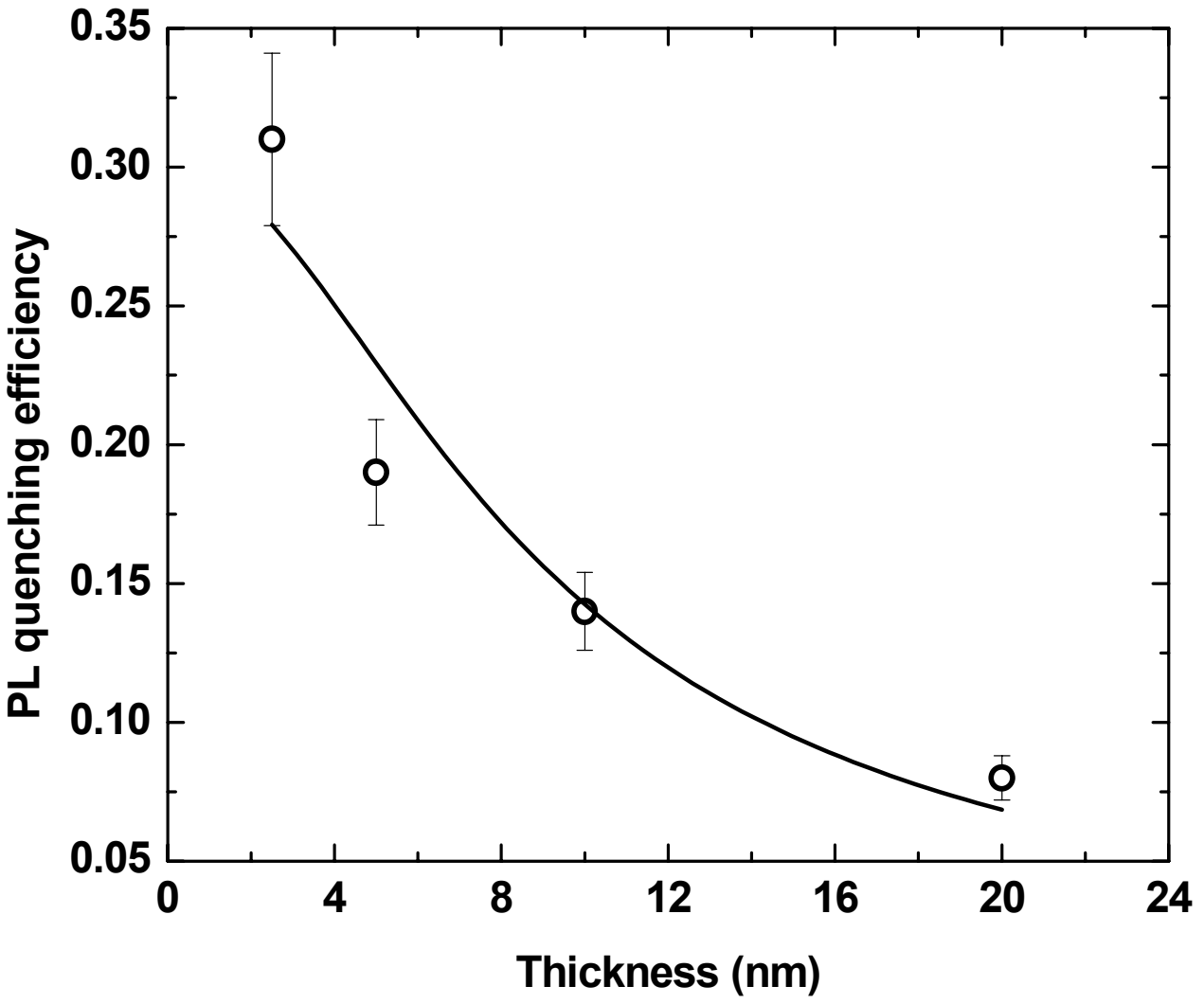

**Fig. 9.** Debdutta Ray et al.

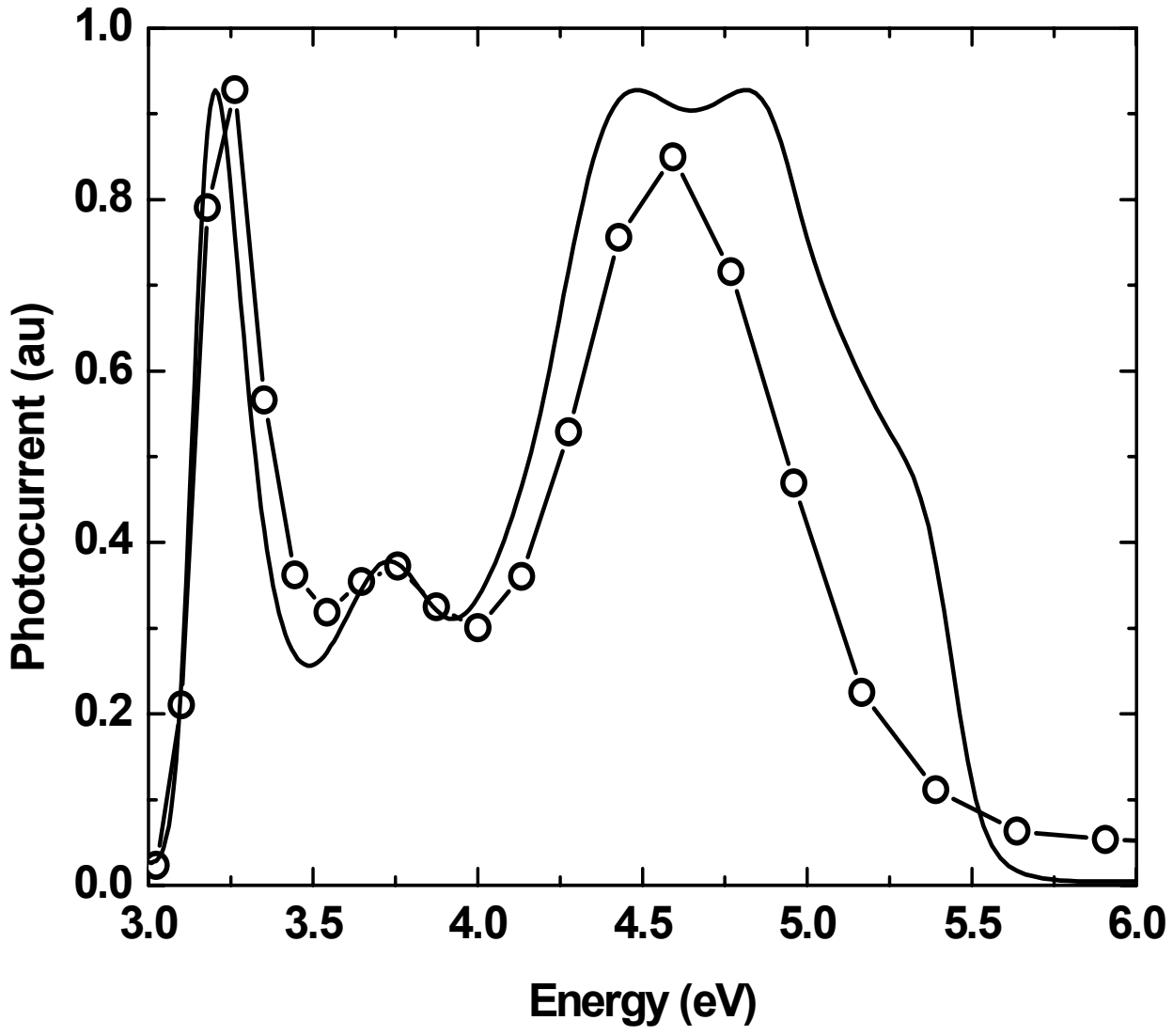

**Fig. 10.** Debdutta Ray et al.

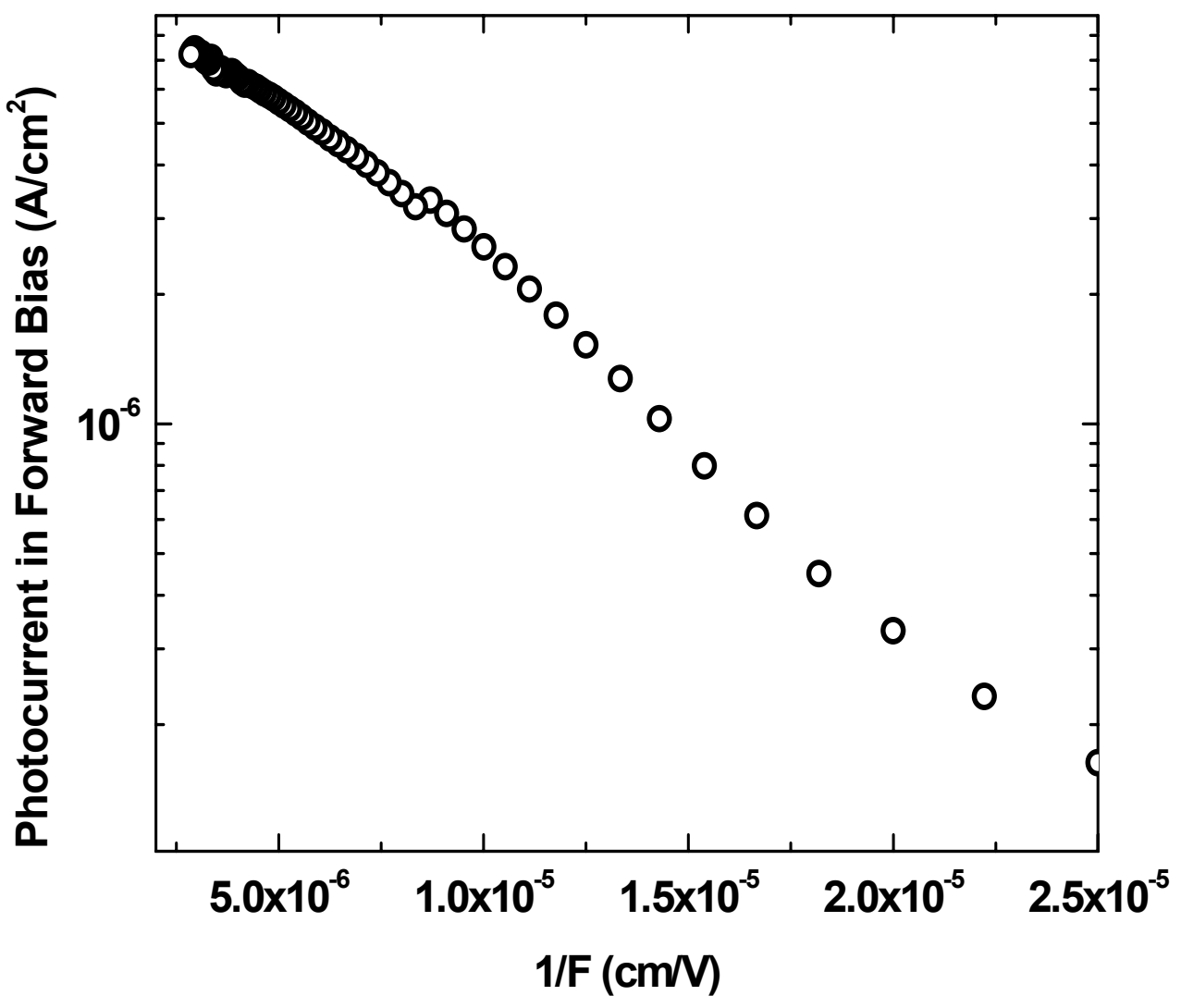

**Fig. 11.** Debdutta Ray et al.